\documentclass[dvips]{article}
\usepackage{icrctc07}

\title{
Constraining Dark Matter hypothesis through Diffuse Source observations with the GLAST-LAT detector
}
\shorttitle{Constraining Dark Matter hypothesis through Diffuse Source observations with the GLAST-LAT detector}
\authors{E. Nuss$^{1}$  on behalf of the GLAST LAT Collaboration}
\shortauthors{G. A. Medina-Tanco and et al}
\afiliations{$^1$ LPTA, Universit\'e Montpellier 2, CNRS/IN2P3, Montpellier, France}
\email{eric.nuss@lpta.in2p3.fr}

\abstract{
The Gamma-Ray Large Area Space Telescope (GLAST), scheduled to be launched in Fall 2007, 
is a next generation high energy gamma-ray observatory. 
The Large Area Telescope (LAT) instrument on-board GLAST
with a wide field of view ($>$ 2 sr), large effective area and 20 MeV to $>$300 GeV energy range, 
will provide excellent opportunity for future Dark Matter studies.
We present an overview of the GLAST Dark Matter and New Physics Working
Group efforts in the study of the LAT capability to detect a gamma-ray flux coming 
from WIMP pair annihilations in diffuse astrophysical sources.
Particular attention will be given to extragalactic diffuse gamma-ray radiation 
and line searches from annihilation into gamma-gamma and/or gamma-Z
final states.
}

\begin{document}
\maketitle

\section{Introduction}

The nature of the Cold Dark Matter (CDM) has been a subject of special interest to 
high-energy physicists, astrophysicists and cosmologists for many years. 
It is probably one of the most outstanding open questions in present day Cosmology.

The existence of Dark Matter is required from galactic scale up to cosmological scale
by a wealth of observations and arguments
such as rotation curve data, cluster dynamics or gravitational arcs.
At the same time, the Cosmic Microwave Background measurements limit the contribution 
from ordinary baryons so that non-baryonic matter has to make up most of the matter in 
the Universe.

Virtually all potential dark matter candidates require physics beyond the standard model
of particle physics.
One of the most widely studied model is 
supersymmetric extensions of the standard model which provides 
a natural candidate for CDM in the form 
of a stable Weakly Interacting Massive Particle (WIMP), the Neutralino.

The mutual annihilation of these WIMPs would yield 
(among a few other indirect signatures like energetic neutrinos, antiprotons 
or positrons) 
many high energy gamma rays ($\ge$1 GeV) that can be well measured in the GLAST LAT. 
WIMP annihilation into $\gamma \gamma$ and or $\gamma Z$ final states would give 
monochromatic gamma rays with an energy $E_\gamma\sim m_\chi$ and $E_\gamma\sim m_\chi\ (1-m_Z^2/4m_\chi^2)$.
Since these gamma rays are monochromatic and have high energy, they would  
constitute a spectacular signature of annihilating dark matter.

The LAT Dark Matter and New Physics Working group has been developing approaches 
for the indirect detection of 
both diffuse and point like \cite{Aldo} source observations with the GLAST-LAT detector \cite{jct}.
We will focus here on diffuse source observations and, as an example,
particular attention will be given to extragalactic diffuse gamma-ray radiation \cite{Conrad} 
and line searches \cite{Edmonds}.

\section{Extragalactic diffuse gamma-ray radiation}

It has been first pointed out by Bergstr\"om et al. \cite{BergCosmo} that 
the integrated signal of WIMP annihilations into high energy photons
over all cosmological dark matter halos at all redshifts 
 might contribute to the extragalactic diffuse gamma-ray radiation. 

The resulting differential energy flux per unit area-time-energy and solid angle on the sky
can be written as \cite{UllioCosmo}:

$${d\Phi_\gamma\over dE_0}={\sigma v\over 8\pi}{c\over H_0}{\bar{\rho}_0^2\over m_\chi^2}\ \times$$
$$
 \int dz(1+z)^3 {\Delta^2(z)\over h(z)} {dN_\gamma(E_0(1+z))\over dE} e^{-\tau (z,E_0)}
$$

where  we can distinguish contributions 
from particle physics, astrophysics and cosmology.

The particle physics contribution lies in the annihilating cross section $\sigma$, the WIMP mass
$m_\chi$, and the differential gamma-ray yield per annihilation :
$${dN_\gamma\over dE}={dN_{cont}\over dE}(E)+b_{\gamma\gamma}\delta(m_\chi-E)$$

Here, $dN_{cont}/dE$ is the mean number of photons due to WIMP annihilations
summed over all relevant channels and gives rise to a continuous energy spectrum. 
This term is reduced to a delta function in the case of monochromatic photons 
as expressed in the second term where $b_{\gamma\gamma}$ is
the branching ratio into $\gamma \gamma$.

As the annihilation rate is proportional to the dark matter density squared
($\bar{\rho}_0$ being the present day mean density),
it depends strongly on how dark matter is distributed on small, galactic and sub-galactic scale.
This question is still a matter of debate however
N-body simulations show that large structures formed by the successive merging of small 
substructures, with smaller objects which are usually more dense \cite{Moore}.
The ``clumpiness''  of dark matter can then significantly boost 
the annihilation signal from cosmological WIMPs.
The quantity $\Delta^2$ describes the average of  over density squared in haloes, as a function of 
redshift.

Since we consider contributions from annihilations at high redshifts,
the extragalactic gamma-ray signal is strongly affected by absorption in 
the inter-galactic medium, especially at high energies.
The term $e^{-\tau (z,E_0)}$ accounts for absorbtion
of gamma-rays along the line of sight where
the absorption is parameterized by the the optical depth $\tau$.
We include the effect of absorption using parameterizations of the optical 
depth as function of both redshift and observed 
energy \cite{Primack}. More recent  calculation of the optical 
depth \cite{Stecker} do not alter our results significantly.

The last contribution comes from cosmological model :  
the Hubble parameter at the present day value $H_0$, and the
dimensionless quantity h(z), 
which depends on the energy content of the universe at a given redshift.
For the fractions of critical density given by matter, vacuum energy and curvature,
we used the results from the WMAP three-year data \cite{Spergel}.

To compute the GLAST LAT sensitivity to Dark Matter contribution to 
the extragalactic diffuse gamma-ray radiation,
we used the Observation Simulator (ObsSim) \cite{ObsSim}.
LAT photons for a generic model of WIMPs annihilating into $b\bar{b}$ 
and into 2 $\gamma$ were generated for different WIMP masses ranging from 50 GeV to 250 GeV.

\begin{figure}
\begin{center}
\includegraphics [height=5.cm,width=0.48\textwidth]{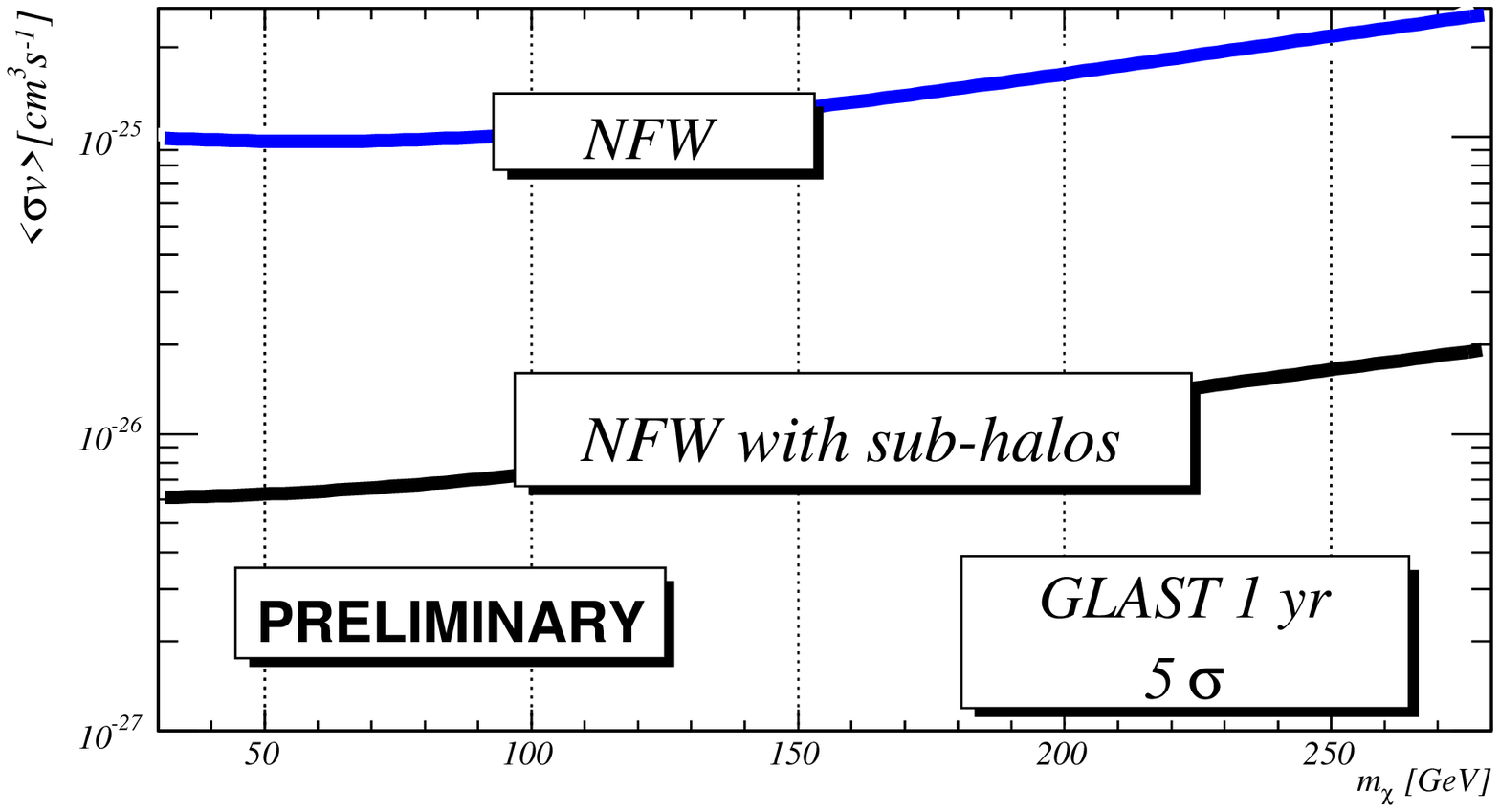}
\end{center}
\caption{
}\label{EGplot}
\end{figure}

Assuming that the background consists of unresolved blazars \cite{UllioCosmo},
a $\chi^2$ analysis was performed, to obtain a 
3 $\sigma$ exclusion curves in the $\sigma v$ vs $m_\chi$ plan for one year of 
simulated data (fig. \ref{EGplot}).

Two different halo profiles have been considered for the normalization :
the Navarro Franck White smooth profile \cite{Navarro} (NFW) and 
NFW profile where we also have included the effect of substructures.
We assumed that substructures constitute 5\% of the mass and have three times the 
concentration parameter of the parent halo.
The concentration parameters, as a function of halo mass, is distributed according to \cite{Bullock}. 

As shown in fig. \ref{EGplot}, depending on the Dark Matter halo model,  
the GLAST LAT would be sensitive to 
annihilating cross-sections from few
$10^{-26}$ up to $10^{-25}\ cm^3 s^{-1}$.

One should note that this estimate neglects contributions of instrumental 
background, uncertainties introduced by the analysis (where point-sources 
and galactic diffuse emission have to be taken into account) and finally 
that the extragalactic background spectrum from astrophysical sources is 
very uncertain, especially at high energies.

\section{Line searches from WIMP annihilation into $\gamma \gamma$ and/or $\gamma Z$ final states}

The detection of monoenergetic photons from mutual annihilation of WIMPs would be a
very distinct ``smoking gun'' signal of Dark Matter content in the Universe.
These processes are however loop suppressed and, in popular SUSY models,
these lines only occur at the $\sim 10^{-4}$ to $\sim 10^{-2}$
branching ratio level, leading to a great challenge in detection.

The LAT Dark Matter and New Physics Working group computed a preliminary estimate 
 of the LAT sensitivity at 5 $\sigma$ above background and 95\% upper limits 
confidence level (ULCL) to these WIMP lines.
Two cases have been investigated, the first being when
the line energy is known (e.g. supplied by SUSY searches at the LHC). 
The second case, when energy is unknown, 
is less constraining as further statistical treatment is needed to account 
for the number of energy bins searched over the range of interest. 

The gamma-ray line signal depends strongly on Dark Matter halo and
on the diffuse background models.
The best signal-to-noise is expected \cite{Stoehr} for an observation 
of a galactic centered broad broken annulus (r $\in$ [$25^o\ ,\ 35^o$]), which 
excludes the region within $10^o$ from the galactic plane.

For this study we estimated the galactic diffuse background in this anulus from
Galprop \cite{Galprop} which is 
based on EGRET and other data in the EGRET energy range.
The high energy range (up to 150 GeV), 
for which the line sensitivity is computed in this proceeding,
has been obtained from
extrapolations provided by the Galprop team.

To estimate the LAT sensitivities to known line energy, we first considered
a point source at high latitude ($l=-76^o$, $b=26^o$) with a narrow Gaussian energy
 distribution ($\sigma/E_0 = 10^{-3}$).
The Instrument Response Functions (IRFs) from the LAT Data Challenge-2 (DC-2)  \cite{DC2}
were used, and LAT resolved lines were simulated with ObsSim 
for a 55 day full sky scanning mode exposure (uniform all-sky coverage for
the LAT in scanning mode).  
Energies of 25, 50, 75, 100, 125, and 150 GeV were considered. 
NB: The DC-2 version of the LAT IRFs cut off
at 200 GeV, the currently available LAT IRF version extends beyond 300 GeV.

The ObsSim results for the all-sky scan were well fit to a double Gaussian distribution, $\Phi_1$ :
$$
\Phi_1(E,E_0,N_T,r,\sigma_1,\sigma_2)=
$$
$${N_T\over\sqrt{2\pi}}
\left[{1-r\over \sigma_1} e^{-{(E-E_0)^2\over 2\sigma_1^2}} + 
      {r\over\sigma_2} e^{-{(E-E_0)^2\over 2\sigma_2^2}} \right]
$$
where $N_T = N_1 + N_2$, and $r = N_2/N_T$ .  

The all-sky diffuse background was generated for 5 years and the flux in the galactic anulus
was fit with an exponential background, $\Phi_2$: 
$$\Phi_2(E;a,b)=a\times e^{E/b} \mbox{ for } E \in [40,200]\ GeV.$$

\begin{figure}
\begin{center}
\includegraphics [height=6cm,width=0.48\textwidth]{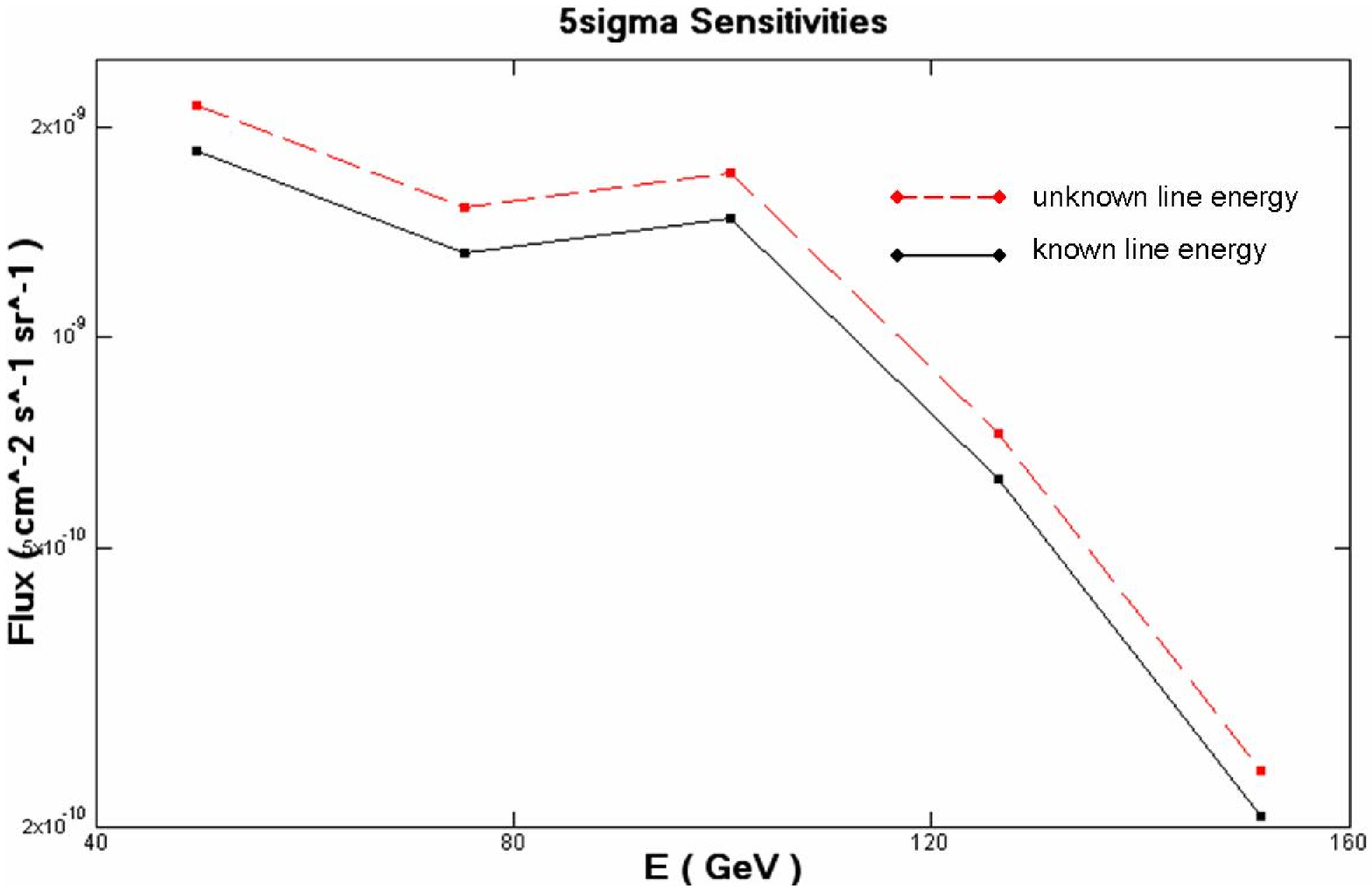}
\end{center}
\caption{
}\label{lines1}
\end{figure}

For each line,
the input background was bootstrapped with a MC signal 1000 times and
fit to $\Phi_1 + \Phi_2$ and $\Phi_2$,
to compute the 5$\sigma$ signal sensitivity 
for 5 years of observations.
This series of 1000 bootstraps was re-run varying only the average 
number of MC signal counts (constant across the series) until 
$<\Delta\chi^2> = \chi^2_{\Phi_1 + \Phi_2}-\chi^2_{\Phi_2}\sim 25$
(corresponding to a 5 $\sigma$ signal over background).  
The average number of signal counts needed at each energy was then converted to a 
sensitivity using average exposures integrated over the annulus (fig \ref{lines1}).  
The LAT detection sensitivity for a line of unknown energy ($E \in [40,150]$ GeV) 
for a 5 $\sigma$ above background signal corresponds to a confidence level of $10^{-7}$.  
To calculate the number of counts needed at each energy in this case we used the 
probability of no detection in a single bin, $q$:
$$q=(1-P)^{1/\#\mbox{ofEnergyBins}}$$
$$={1\over\sqrt{2\pi}}\int_{-\infty}^{\# of\sigma} e^{-x^2/2}\ dx,$$
where $P = 10^{-7}$
is the probability of detecting a signal greater than $\# of\sigma$, in one or more bins.  
A bin width of $\sigma_{fwhm}/E = 8\%$ was used based on the FWHM energy resolution of the IRFs.  
The error in the signal was estimated as $\sigma_{signal}=\sqrt{2 B}$ where $B$ is the number
of background counts in a bin centered on $E_0$ with width $\sigma_{fwhm}$.  
The $\# of\sigma$ signal error was converted to flux for calculation of the 
unknown line energy (fig. \ref{lines1}).

\begin{figure}
\begin{center}
\includegraphics [height=6cm,width=0.48\textwidth]{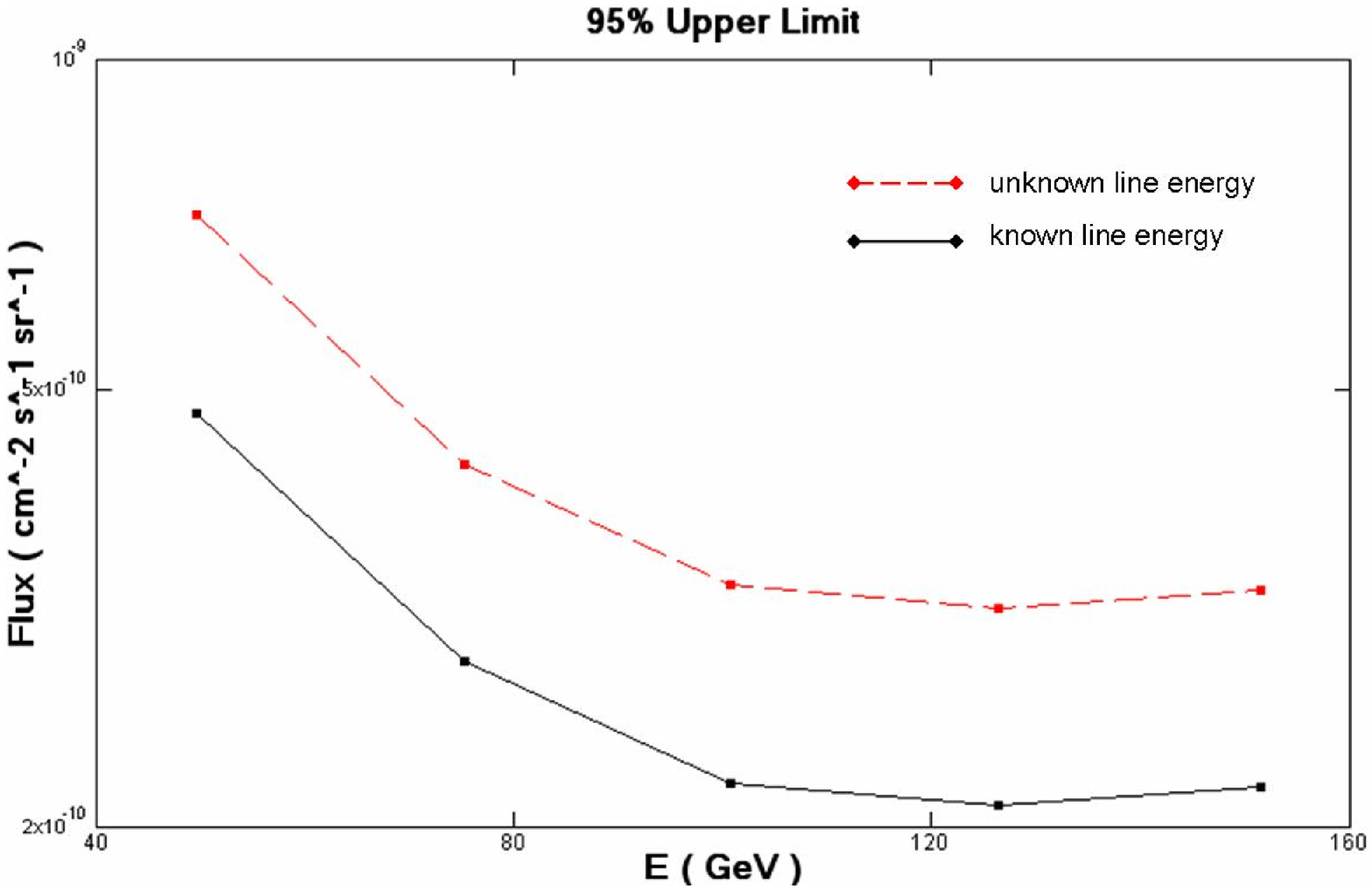}
\end{center}
\caption{
}\label{lines2}
\end{figure}

The 5 year 95\% CLUL sensitivity for known line energies was obtained similarly to 
the 5$\sigma$ case by bootstrapping the ObsSim diffuse background and fitting $\Phi_1 + \Phi_2$ with 
no MC generated signal.  
After the calculation of $N_T \pm\sigma_{NT}$, the 95\% CLUL was obtained from 1.64 $\sigma_{NT}$
and sensitivities calculated as in the $5\sigma$ case with P=.05 (fig \ref{lines2}).  
The case of unknown line energy ($E \in [40,150]$ GeV) 95\% CLUL was calculated as in the 5$\sigma$ case 
with P = .05 ( fig \ref{lines2}).  

\section{Conclusions}

We presented the GLAST Dark Matter and New Physics Working
Group efforts 
on indirect detection  of Dark Matter through 
diffuse source observations.
As an example, preliminary results on  
extragalactic diffuse gamma-ray radiation
and line searches
illustrated that
the GLAST-LAT detector
will provide excellent 
high energy gamma-ray observations for
 future Dark Matter searches.
\bibliography{icrc0625}
\bibliographystyle{plain}

\end{document}